\documentclass[aps,pre,twocolumn,floatfix]{revtex4}
\usepackage{graphicx}
\usepackage{pst-all}
\usepackage{color}
\usepackage{amsmath}
\newcommand{\comment}[1]{}

\begin{document}
\title{Evolution of a population of random Boolean networks}
\author{Tamara Mihaljev and Barbara Drossel}
\affiliation{Institut f\"ur Festk\"orperphysik,  TU Darmstadt,
Hochschulstra\ss e 6, 64289 Darmstadt, Germany }
\date{\today}
\begin{abstract}
\end{abstract}
 \begin{abstract}
We investigate the evolution of populations of random Boolean networks
under selection for robustness of the dynamics with respect to the
perturbation of the state of a node. The fitness landscape contains a huge
plateau of maximum fitness that spans the entire network space.  When
selection is so strong that it dominates over drift, the evolutionary
process is accompanied by a slow increase in the mean connectivity and
a slow decrease in the mean fitness.  Populations evolved with higher
mutation rates show a higher robustness under mutations.  This means
that even though all the evolved populations exist close to the plateau of
maximum fitness, they end up in different regions of network space.
\end{abstract}
\maketitle

\section{Introduction}
\label{intro}

The Neodarwinian view of biological evolution considers random
mutations and natural selection as the main shaping forces of
organisms. Mutations act on the genotype, while selection acts on the
phenotype. For this reason, the relation between genotype and fitness
is very complicated and far from fully understood. Mathematical models
of biological evolution \cite{BarbaraRevEvo} often contain a direct mapping
of the genotype on the fitness. The ``fitness landscape'' may be
smooth and single-peaked or random and rugged, or the fitness is taken
as the additive contribution of the alleles at several loci. However,
the ``real'' fitness landscape might have completely different
properties. For this reason, it is important to investigate models
that do not make a direct mapping of the genotype to the fitness, but
that determine the fitness from some ``trait'' that is related in a
nontrivial way to the genotype.

The most famous example of such models are based on RNA. The genotype
is the RNA sequence, while the phenotype is the two-dimensional
fold. When fitness is based on some desired fold, it is found that the
fitness landscape contains a huge plateau of high fitness that spans
the genotype space \cite{fon93,sel94}. The same feature is displayed
by the fitness landscape that is based on the three-dimensional fold
of proteins, with the genotype being the nucleotide sequence of the
corresponding gene \cite{porto_collaboration}.

However, most traits of an organism result from the interaction of
many genes. For instance, most genes are very similar in different
higher organisms, but they differ in the way they are regulated and in
the temporal expression pattern during embryonic development.  This
feature is captured in models for gene regulation networks, the
simplest of which is the random Boolean model introduced 
in 1969 by Stuart Kauffman 
\cite{kau1969a,kau1969b}. In this simple model, each gene $i$ can be in
two different states, that is the state $\sigma_i$ can be either
``on'' (1) or ``off'' (0). This means that the gene is either
expressed or not. Each gene is represented by a node and each
interaction by a directed connection between two nodes.  Each node $i$
receives input from $K_i$ randomly chosen other nodes, and its state
at time step $t$ is a function of the states at time step $t-1$ of
the nodes connected to it,
\begin{equation}
\sigma_i(t) = f_i[\sigma_{i_1}(t-1), \sigma_{i_2}(t-1), ...,
  \sigma_{i_{K_i}}(t-1)] \label{update}
\end{equation}
Starting from any of the $2^N$ possible states $\vec \sigma =
 \{\sigma_1, ..., \sigma_N\}$ the network eventually settles on a
 periodic attractor. Usually, there are different attractors with
 different basins of attraction (i.e., the fraction of states leading
 to and lying on the attractor) and attractor lengths (number of
 states the attractor consists of). Thus, the dynamical behaviour of a
 Boolean network is its phenotype, while the genotype is specified by
 the logical functions and the connections between the nodes.

Several publications study the evolution of such Boolean
networks. Mutations are performed by changing the connections or
functions. In addition, several investigations include recombination
of the parental genotypes in the evolutionary simulations
\cite{frank99population,wagnerPlasticity,sevimEvol07}. 
 In \cite{aldana07genDuplication} the effect of
gene duplications was additionally studied. 

Selection is based on some dynamical property of the
networks. In \cite{Kauffman86,Stern99,oikonomou06} and \cite{Lemke01},
the fitness is given by the distance of an attractor to a predetermined target
pattern.  In \cite{Bornholdt98,Bornholdt00}, the selection criterion
requires that the daughter network reaches the same attractor as the
mother network when both networks are initialised in the same randomly
chosen state.  In \cite{Paczuski00,Luque01,BornholdtRohlf00,Liu06},
mutations are targeted to those nodes that display a certain type of
behaviour.

In \cite{Agnes07}, the fitness criterion is robustness of the dynamics
under small perturbations. Robustness is of great importance in
biology as a cell has to maintain its biological functions to survive
and pass on its genetic material under variations for example of the
concentration of proteins in the cell or of the nutrient level. In
\cite{Agnes07}, evolution was simulated by means of a so-called adaptive
walk. This is a hill climbing process that leads to a local maximum in
the fitness landscape and thus can yield insight in the fitness
landscape of a system. The main finding was that the maximum possible
fitness value is always reached after a few mutations during this process, 
and that there is a huge plateau with this fitness value that spans the 
network configuration space. 

In this paper, we study the evolution of an entire population of
networks under the mutation and selection rules employed in
\cite{Agnes07}. We investigate the fitness and the diversity of the
population as a function of time, as well as the topological
properties of the evolved networks as function of the mutation rate
and the selection pressure. The most interesting finding is
that, while the population quickly reaches the plateau of high
fitness, the network topology undergoes a very slow change towards
higher connectivity, while at the same time the mean fitness of the
population decreases slightly. 

The outline of this paper is as follows: In section \ref{model}, we
present the rules of our evolutionary model. In section \ref{pr0}, we
investigate the evolutionary process in the absence of selection,
where random mutations and genetic drift are the only shaping
forces. In section \ref{prinfty}, we study the opposite case  of
very strong selection, where only the networks with the highest  
fitness value become parents of the networks in the next
generation. In section \ref{generalpr}, we then study the general case
of finite selection pressure. Section \ref{conclusion} summarises and
discusses our findings.


\section{Model}\label{model}

A population of $P$ networks with $N$ nodes each is evolved by repeatedly
replacing the entire population with a daughter population. Each
individual in the daughter population is obtained by choosing an
individual from the parent population to be its mother with a
probability that depends on the mother's fitness. The daughter is
a copy of the mother, but it receives one mutation
with a certain probability $\mu$. 

The initial population is generated by connecting the nodes of each
network at random, with $K_{ini}=3$ inputs per node, and with the update
function of each node chosen at random from the set of canalyzing
functions used by Moreira and Amaral \cite{mor2005}. Thus, the
function at a node is determined by choosing one of the input nodes as
the canalyzing input. Its canalyzing value and the associated output
value each are 0 or 1 with probability 1/2. When the input node is not
on its canalyzing value, the output is a random Boolean function that
depends on the remaining variables and is generated by choosing with
same probability $0$ or $1$ as output for every combination of the 
remaining input variables. The reason for choosing  $K_{ini}=3$  is
that this is the critical value for this class of networks. The
initial networks are therefore neither completely frozen, nor are they
chaotic (in the sense that neighboring initial states diverge
exponentially fast). 

The fitness of a network is determined by the following rule: First,
the network is initialised in a random state and is updated according
to equation (\ref{update}) until it reaches an attractor. Then the
value of each node is flipped one after the other, and it is counted
how often the network returns to the same attractor. This can happen
at most $N$ times. The fitness value $f$ is the percentage of times the
dynamics return to the given attractor after flipping a node. The
weight with which an individual $i$ is chosen to be the mother of a
given individual of the next generation is 
\begin{equation}
w_i = \frac{e^{pf_i}}{\sum_{j=1}^N  e^{pf_j}} \, ,
\end{equation}
where we call $p$ the selection pressure. In addition to $P$ and $N$ and
$\mu$, this is the fourth parameter that was varied in the simulations. 

Four different mutations can occur, each with the same probability:
\begin{enumerate}
\item A connection is added.
\item A connection is deleted.
\item A connection is redirected.
\item The canalyzing part of the function is changed.
\end{enumerate}

When a network is to undergo a mutation, first the type of mutation is
chosen.  Then, a node is picked at random to receive this mutation. If
the mutation cannot be performed at this node (for instance, only the
mutation of adding a link can be done at a node with zero inputs)
another node is selected at random to receive the mutation. Due to
computational restrictions, we imposed the rule that nodes with 10
inputs cannot receive an additional link, and therefore $K_{max} =
10$.

When a connection is added or deleted, the Boolean function of the
node has to be changed. This is done by choosing anew the random
Boolean function that depends on the non-canalyzing variables.  If the
canalyzing input is removed by the mutation, another node takes its
role.

A connection is redirected by changing at random the origin (source)
of one incoming link of the node which is receiving the mutation. 
Finally, when changing the canalyzing function of the node, the value that
canalyzes it and the associated output value are assigned to the node anew.

\section{Evolution without selection}\label{pr0}

In the absence of selection, each network has the same probability $1/P$ to
become the mother of a given daughter network. The average  number of
generations back to the last common ancestor of two networks is
therefore $P$. With probability $\mu$, a daughter network receives a
mutation, and therefore two randomly chosen networks of the population
differ by $2P\mu$ mutations on average. 

With probability 1/2, the mutation consists in the addition or
deletion of a link. This means that the total number of links in the
network performs a random walk in time, with probability $\mu/2$ for a
nonzero step.  Even though the number of inputs is initially
$K_{ini}=3$ for each node, it changes during time, and the
distribution of the number of inputs becomes eventually stationary.
The probability that the number of inputs of a node changes during a
given step depends on the number of nodes with 0 and 10 inputs.

\begin{figure}
\includegraphics*[width=0.4\textwidth]{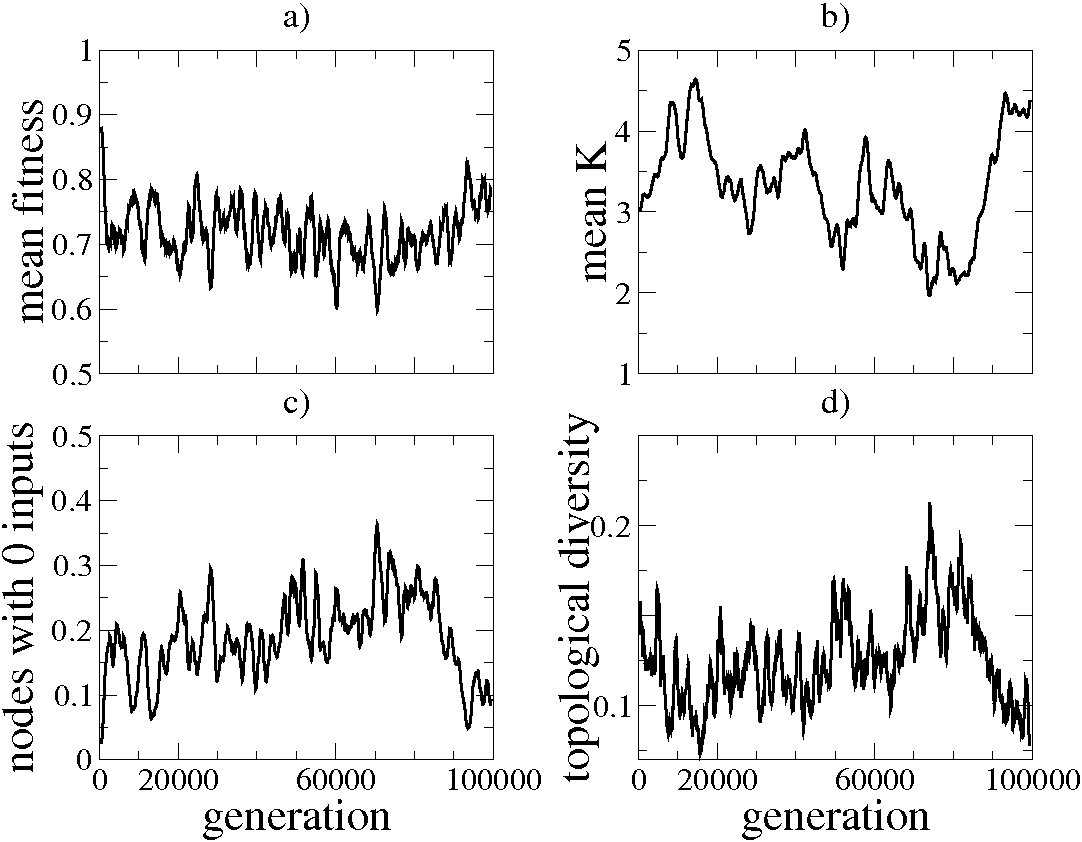}
\caption{
Evolution of the  a) mean fitness, b) mean number of inputs per node, c) mean
number of nodes with zero inputs in a network,  d) proportion of different
links in two randomly chosen networks, in the population of 50 networks of 50
nodes each, evolved with the mutation rate 0.5 and no selection pressure. 
}
\label{Pr0}
\end{figure}

These simple considerations are very useful when interpreting the
simulation data. Figure \ref{Pr0} shows results of simulation run for
100000 generations of a population with $N=P=50$ and with a mutation
rate $\mu=0.5$.  We evaluated the mean fitness of the population, the
mean number of inputs per node, the proportion of nodes with 0 inputs,
and the topological diversity of the population.  The topological
diversity of the population is the average number of links of a
network that is not shared by a randomly chosen other network, divided
by the mean number of links per network.

The data were smoothened by averaging each data point over 1000
generations, otherwise the data are so noisy that variations on larger time
scales are hard to see.

From these data, one can draw the following conclusions:
\begin{itemize}
\item The mean fitness and the mean number of inputs per node show
  large fluctuations over time. This is due to the fact that the last
  common ancestor of the population is not much more than 50
  generations back, which means that the networks in the population
  are strongly correlated. 

\item The initial fitness of the population is higher than that at
  later times. This must be due to the changes occurring in network
  structure, i.e. to the distribution of the number of inputs becoming
  broader. In particular, nodes with zero inputs decrease the fitness
  (see next point).

\item There is an anticorrelation between the fitness and the
  proportion of nodes with 0 inputs. Clearly, a node with 0 inputs
  does not return to its initial state after a perturbation. If all
  other nodes did return to the same attractor after a perturbation,
  the fitness would be identical to the proportion of nodes with at
  least 1 input. 

\item There is an anticorrelation between the mean number of inputs
  and the topological diversity. This is due to the fact that the
  probability that a given link is mutated becomes smaller when there
  are more links.

\end{itemize}

\begin{figure}
\includegraphics*[width=0.4\textwidth]{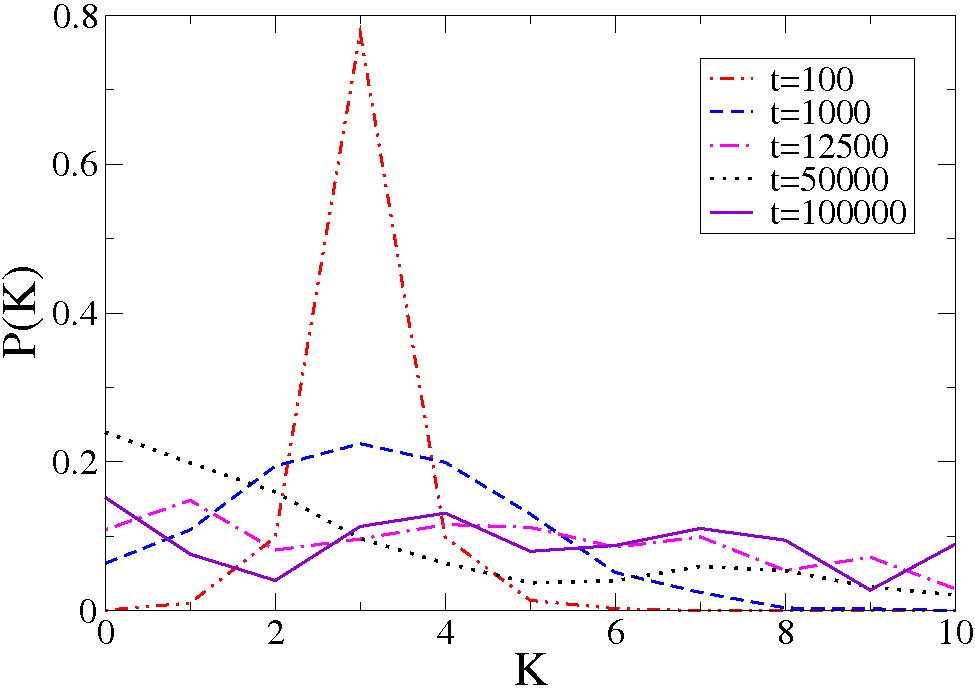}
\caption{ Snapshots of the input distribution at different times
  during the evolution without selection pressure. The input
  distribution is changing from a delta peak to a broad distribution
  with an increased number of nodes with zero inputs.  }
\label{InPr0}
\end{figure}

\begin{figure}
\includegraphics*[width=0.4\textwidth]{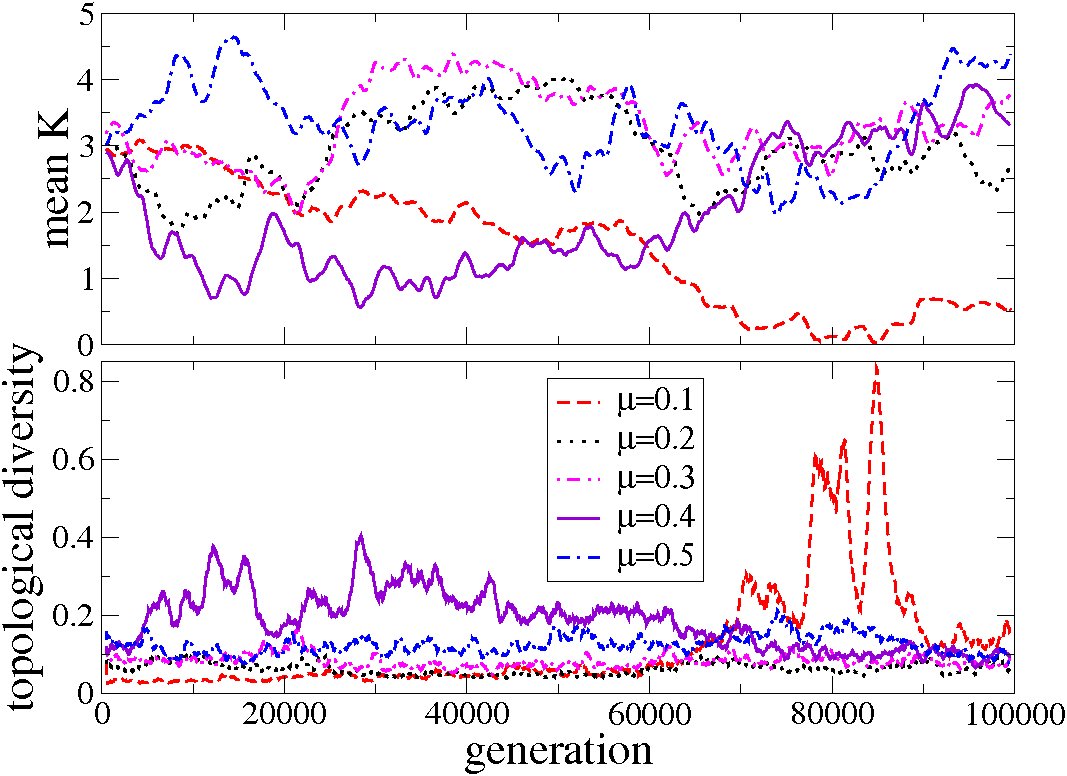}
\caption{ Evolution of the mean value of the number of inputs per node
  (upper figure) and of the topological diversity in the population
  (lower figure) under no selection pressure and with different
  mutation rates.  }
\label{DifMDiv+Kp0}
\end{figure}

The change in the distribution of the number of inputs is illustrated
in Fig. \ref{InPr0}. It is becoming broader, with more nodes with
zero inputs,  which can temporarily even become the
dominant type of nodes in the network. (An example for this is given below
in one of the cases illustrated by Fig.~\ref{DifMDiv+Kp0}.)

Let us now apply analytical considerations in order to estimate the
topological diversity of the population.  Let $T$ be the average time
(in terms of the number of generations) to the last common ancestor of
two randomly chosen networks in the population. With no selection
pressure, we have $T=P$, but with selection this time becomes shorter.
Since their last common ancestor, each network received in each
generation a mutation with a probability $\mu$.  Therefore the two
networks together have received on average $2T\mu$ mutations.  If the
effect of each of them is different and if there are no back
mutations, two randomly chosen networks in the population differ on
average by $2T\mu$ links and functions. When evaluating the
topological diversity, we consider only links. Since three out of
four mutations affect links, we expect that two randomly chosen
networks in the population have received together $3T\mu/2$ mutations
of links. All these mutations affect different links only if $3T\mu/2$
is small compared to the total number of links of a network, $NK$.  In
this case, the topological diversity should be close to
$\dfrac{3T\mu}{2NK}$. Otherwise, it is smaller, since two mutations
may affect the same link. 

Since $K$ fluctuates strongly with time, the topological diversity
fluctuates also and should behave approximately as $1/K$. In figure
\ref{Pr0}d) we have already seen this anticorrelation. For small $K$,
the number of links that are different in the two networks is much
smaller than the number of the link mutations they received, and the
topological diversity does not become as large as suggested by our
simple estimate. This effect is nicely demonstrated by Figure
\ref{DifMDiv+Kp0}, which shows the mean value of $K$ and the
topological diversity as function of time for five simulation runs
with different mutation rates $\mu$.

\begin{figure}

\includegraphics*[width=0.4\textwidth]{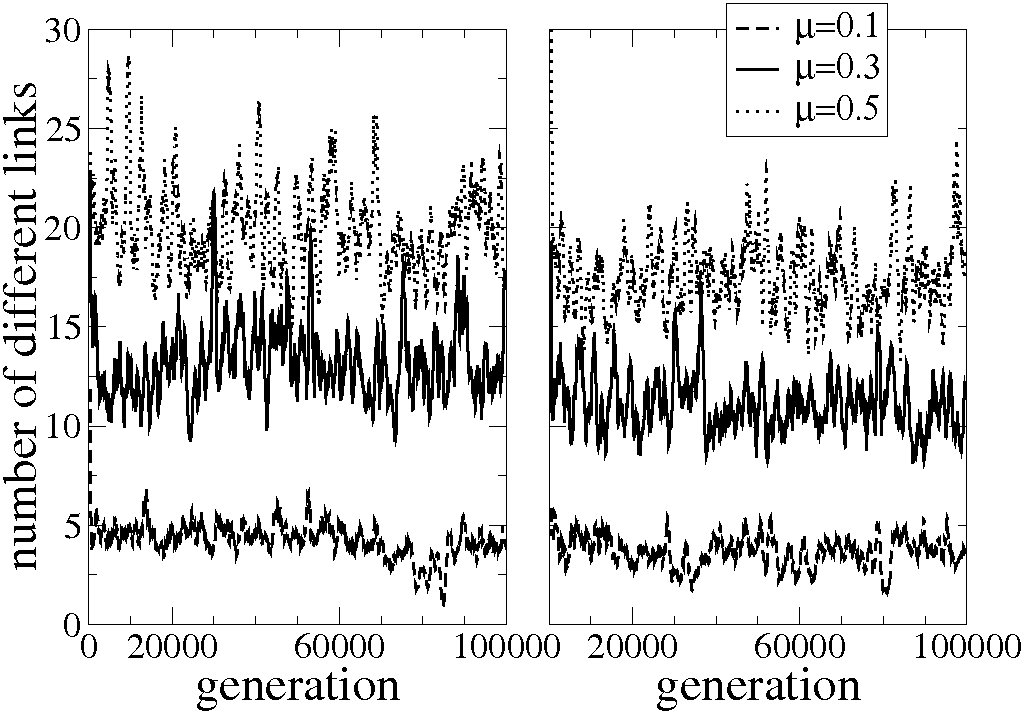}
\caption{ Average number of links by which two networks differ, as
function of time and for different mutation rates. The selection pressure is
zero in the left graph and 100 in the right graph. }

\label{DivMutdifMp0+100}
\end{figure}

Figure \ref{DivMutdifMp0+100} left shows the number of links by which two
networks differ on average, which is the topological diversity
multiplied by $NK$.  Our above simple estimate gives link numbers of
7.5, 22.5 and 37.5 for the mutation rates shown in the figure. These numbers are
upper bounds, and one can see that for larger mutation rates and for smaller $K$
values the data are farther below these bounds, since multiple mutations of the
same link become more frequent.

We investigated also the influence of the size and of the number of
networks in the population on the properties of evolved populations by
setting $N$ and/or $P$ to 30. Most of these properties depend strongly on
the mean number of inputs per node, which performs a random walk and shows
therefore large fluctuations. For this reason, we could not see a clear trend with $N$
or $P$ in the simulation data, although one can expect that 
the fitness should not depend on $N$ or $P$ and that the fluctuations
should decrease with increasing $P$. The diversity should change with $N$ and $P$ as
$P/N$, as predicted by analytical estimation earlier in this section.

\section{Evolution with very strong selection}\label{prinfty}

Next, we consider the opposite case of very strong selection. In this
case, only the networks with the highest fitness value in the
population become parents. Now the properties of the fitness landscape
play an essential role at determining the evolution of the population.
If there were isolated peaks in the fitness landscape, the entire
population would perform a hillclimbing process. The fittest
individual of the parent population would be the mother of all
individuals in the next generation, which would differ from it by at
most one mutation. If one of these mutations did lead to a higher
fitness, all individuals of the following generation were descendants
of the carrier of that mutation. The process would end at a local peak
of the fitness landscape, from where no mutation is possible that
increases or retains the fitness value.

However, when the fitness landscape has a plateau with maximum fitness
that spans the entire network configuration space, the population can
contain several individuals with fitness 1, and mutations can generate
other genotypes with fitness 1. In our simulations, already the initial
population may contain an individual with fitness 1, so that
the mean fitness can be close to 1 already after one generation.

\begin{figure}
\includegraphics*[width=0.4\textwidth]{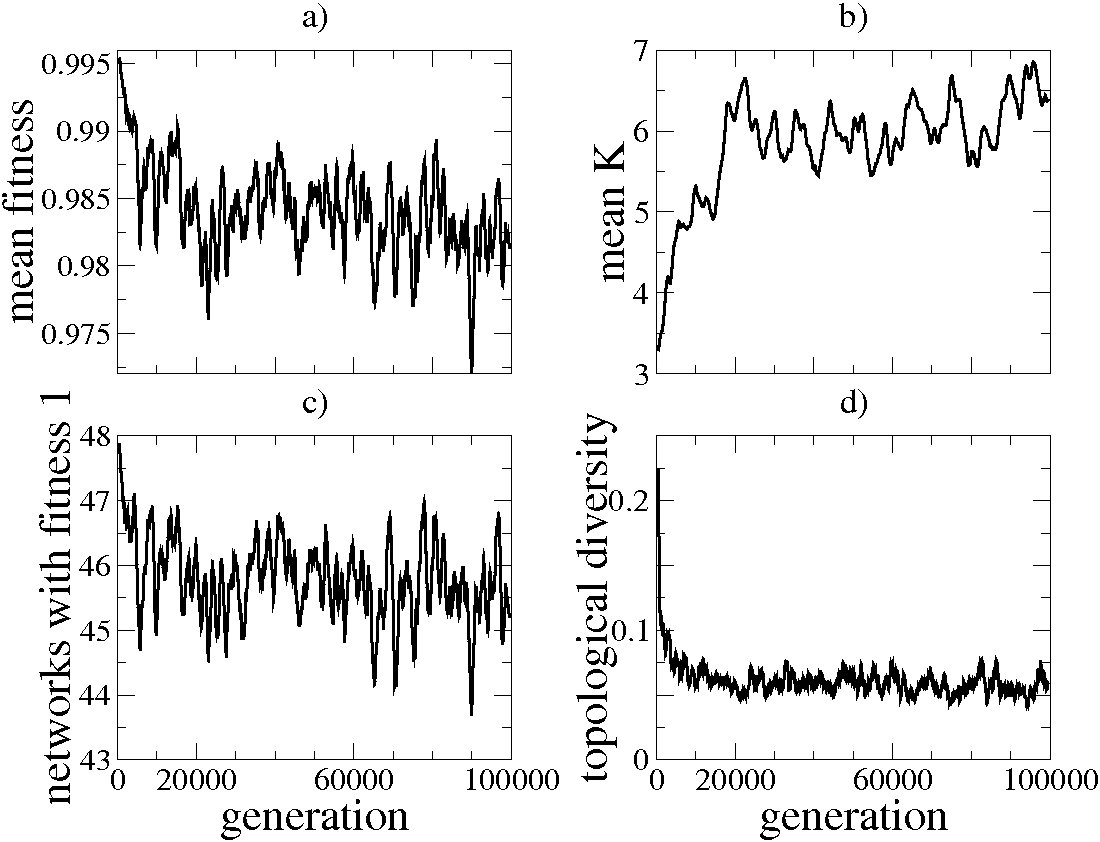}
\caption{Evolution of the  a) mean fitness, b) mean number of inputs per node,
c) mean number of networks with fitness 1,  d) proportion of different
links in two randomly chosen networks, in the population of 50 networks of 50
nodes each, evolved with the mutation rate 0.5 and high selection pressure. 
}
\label{Pr100}
\end{figure}

Figure \ref{Pr100} shows results of computer simulations for 100000
generations of a population with $N=P=50$ and with a mutation rate
$\mu=0.5$.  The parameters are the same as in Figure \ref{Pr0}, and
each data point represents again an average over 1000 generations.  We
evaluated the same quantities as in the first simulation, except for
the number of nodes with 0 inputs, since these do not occur any more.
Instead, we show the number of networks with fitness 1 in the
population. The proportion of networks with a fitness smaller than 1
must be identical to $\mu$ times the probability that a mutation
decreases the fitness of a network with fitness 1.

From these data, one can draw the following conclusions:
\begin{itemize}
\item The fitness of the population decreases slowly with time. Since 
all networks with a fitness smaller than 1 must be daughters of
networks with fitness 1, this means that at later times the average
fitness decrease due to a mutation must be larger.  
\item The mean number of networks with fitness 1 in the population
  decreases slowly with time. This means that the probability that a
  mutation decreases the fitness of the network with fitness 1 is
  larger at later times. 
\item The mean number of inputs per node increases slowly but
  steadily. This was already found in the adaptive walk simulations in
  \cite{Agnes07}. This means that mutations that preserve the maximum
  fitness are more likely to occur when a link is added than when a
  link is removed.
\end{itemize}

\begin{figure}
\includegraphics*[width=0.4\textwidth]{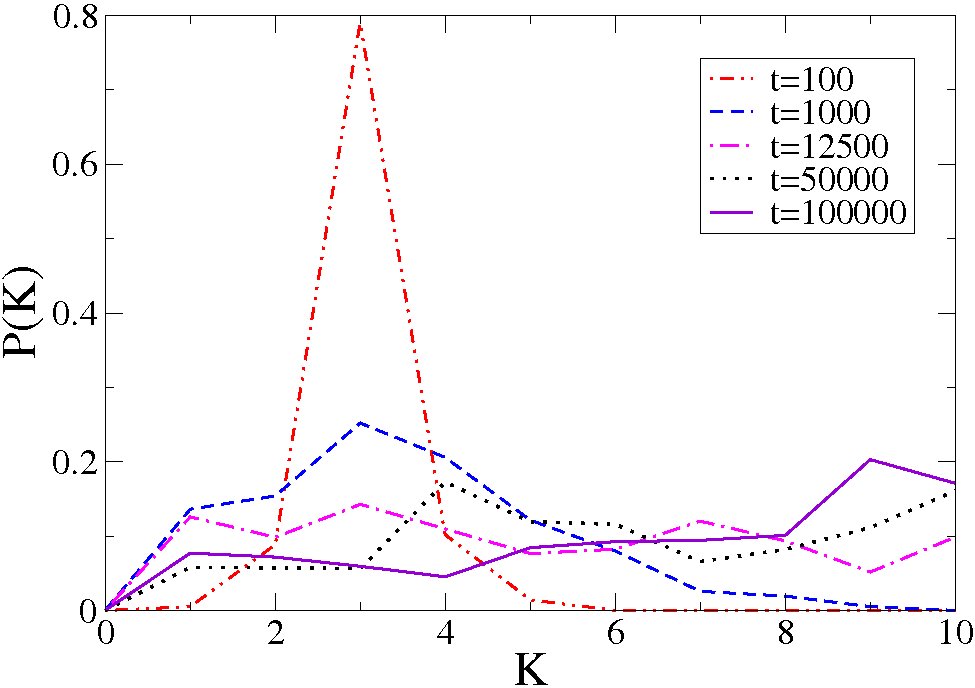}
\caption{
Snapshots of the input distribution at different times during the evolution. The
input distribution is changing from a delta peak to a broad distribution. 
}
\label{InPr100}
\end{figure}

The change in the distribution of the number of inputs is illustrated
in Fig. \ref{InPr100}. As in the situation without
selection, it is becoming broader, but now there are more nodes with
higher $K$ and less with smaller $K$. One reason for this is that
nodes with zero inputs decrease the fitness, and therefore evolution
drives the population into regions in configuration space where such
nodes are unlikely to occur. 

\begin{figure}
\includegraphics*[width=0.4\textwidth]{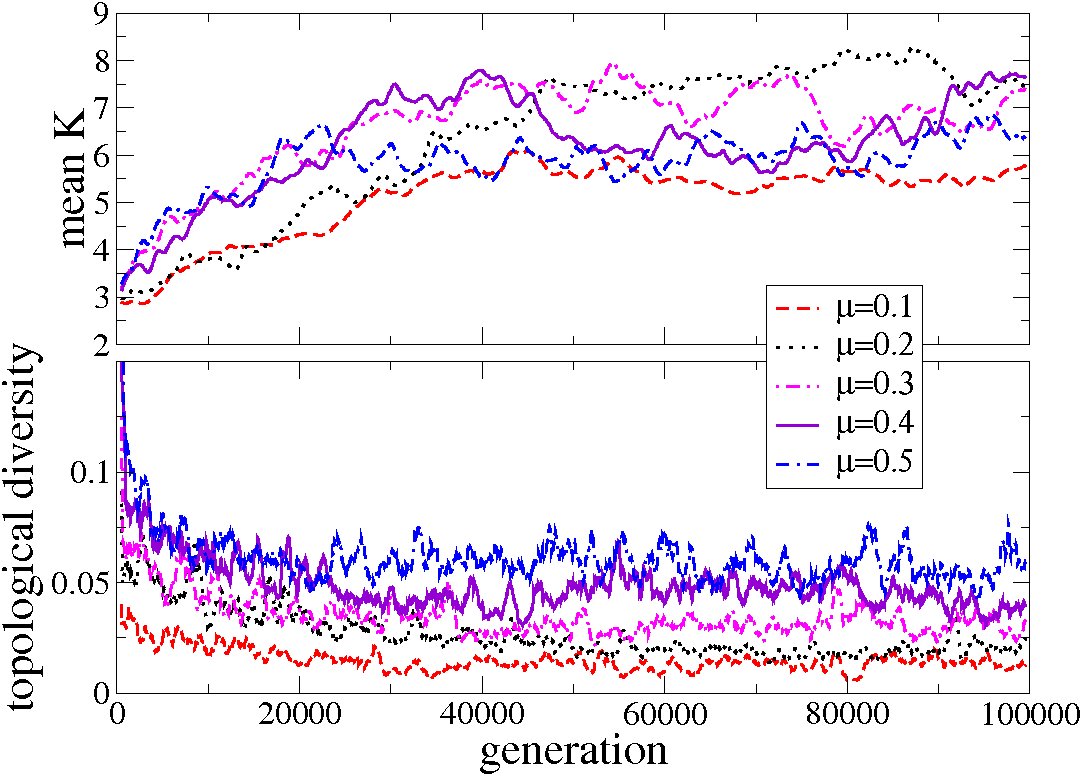}
\caption{ Evolution of the mean value of the number of inputs per node
  (upper figure) and of the topological diversity in the population
  (lower figure) under strong selection and with different
  mutation rates. }
\label{DifMDiv+Kp100}
\end{figure}

Let us now estimate the topological diversity using the arguments from
the beginning of Section \ref{pr0}.  The number of networks with
fitness 1 in the population is defining the effective population size
$P'$, since only these networks can become parents.  From
Fig.~\ref{Pr100}c), we see that this number is around 46 for the
parameter values used in this simulation. Correspondingly, the data
for the number of links that differ between two randomly chosen
networks (Fig.~\ref{DivMutdifMp0+100} right) are slightly lower
compared to the data obtained with zero selection pressure. The
estimated upper bound for the topological diversity is now
$(3P'\mu)/(2NK)$, with deviations from this bound being again larger
for smaller $K$ and larger $\mu$.  In Figure \ref{DifMDiv+Kp100}, it
can clearly be seen that the topological diversity depends on the mean
$K$ value.

We explored in more detail the effect of mutations on networks with
fitness 1. The probability that a mutation does not decrease fitness
is identical to the proportion of networks with fitness 1, shown in
Figure \ref{Pr100}c). This is because selection pressure is so high
that only networks with fitness 1 become parents of the networks in
the next generation, each of which then receives a mutation with
probability $\mu$. We call a mutation that does not decrease the fitness
"neutral".

\begin{figure}
\includegraphics*[width=0.4\textwidth]{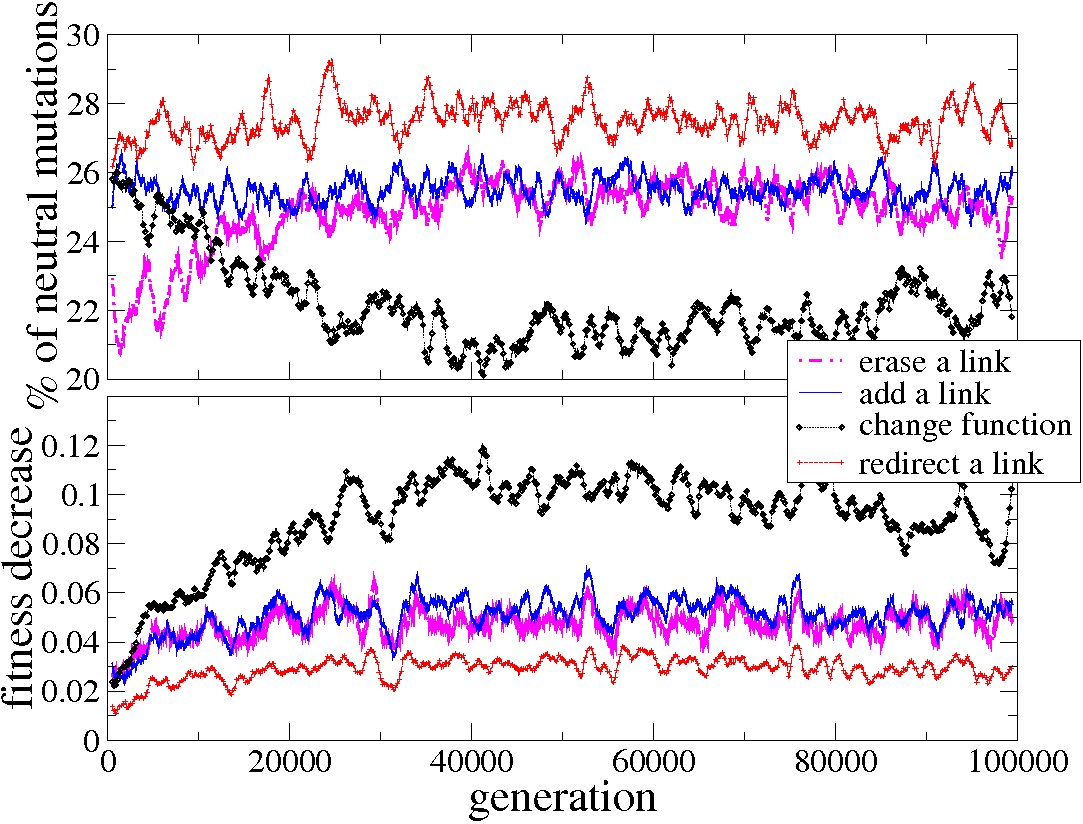}
\caption{ Evolution of the proportion of neutral mutations of different types
  (upper figure) and of the mean fitness decrease per non-neutral mutation
 (lower figure) under strong selection pressure and
with a mutation rate $\mu=0.5$. }
\label{mutations}
\end{figure}

Figure \ref{mutations} (upper panel) shows the proportion of the four
different types of mutations among these neutral mutations, again for
$\mu=0.5$. All four types of mutations were chosen equally often,
however, the proportion of neutral mutations is different for the four
mutation types.  The most frequent neutral mutation is the redirection
of a link.  This means that networks with fitness 1 are most robust
(in the sense that their fitness is not decreased) to this type of
mutations. The least frequent neutral mutation is initially the
deletion of a link; at later times the change of a function is least
frequent. The combined contribution of these two types of mutations to the neutral
mutations is approximately constant in time. The frequency of mutations
that add a link is also approximately constant. This implies that the
slow increase of the mean $K$ value is not due to a beneficial effect
of mutations that add links, but due to the fact that erasing a link
decreases the fitness more often than the addition of a link, in
particular in the beginning of the simulation, where also the largest
increase of $K$ can be seen. At later times, deletions and additions
are equally frequent among the neutral mutations. 
The increase of $K$ explains why mutations that change the
canalyzing function become less frequent among neutral mutations. When $K$ is
larger, such a change affects more nodes on an average. 

In Figure \ref{mutations} (lower panel), we show the
amount by which fitness decreases due to a non-neutral mutation. 
This amount increases with time for all four types of mutations. 
This must be due to the fact that the perturbation of one node affects
more nodes when $K$ is larger, and therefore a mutation affects also
more nodes. For the same reason, mutations that change the canalyzing
function lead to a larger fitness decrease than other mutations. The
redirection of a link leads to the smallest fitness decrease because
it does not involve changes in the update functions. 

\begin{figure}
\includegraphics*[width=0.4\textwidth]{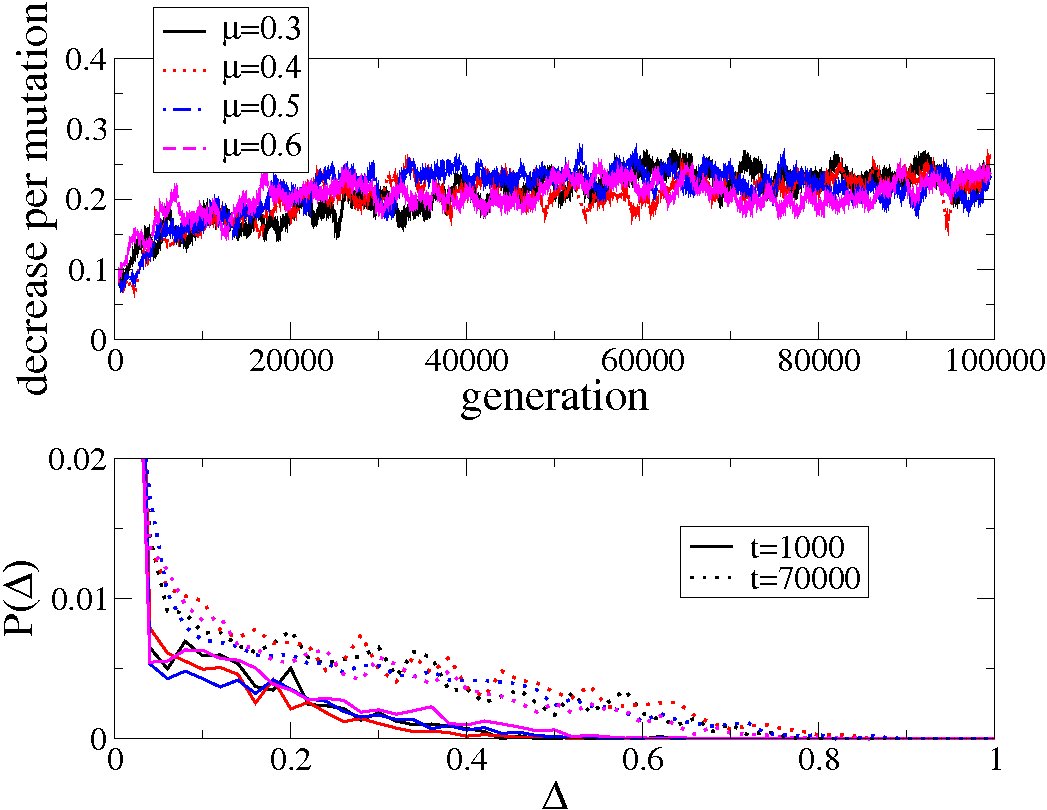}
\caption{Upper graph: Fitness decrease per (non-neutral) mutation as
  function of time, for four different mutation rates.  Lower graph:
  Probability distribution of the fitness decrease $\Delta$ (including
  zero decrease) due to a mutation, evaluated at two different times
  during evolution. }
\label{decrease+distrib}
\end{figure}

Next, we investigated the influence of the mutation rate on these
results.  Figure \ref{decrease+distrib} (top) shows the fitness
decrease per mutation for different mutation rates.  Here, we now do
not discriminate between different types of mutations. The mean
fitness decrease per non-neutral mutation appears to be independent of
the mutation rate with which the networks were evolved.

The lower graph of Figure \ref{decrease+distrib} shows the probability
distribution of the fitness decrease per mutation at an early and at a
late time. Here, neutral mutations, which lead to a fitness decrease
of 0, are also included.  After approximately 30000 generations, the
probability distribution of the fitness decrease reaches a stationary
shape. This shape does not depend on the mutation rate with which the
networks were evolved.  Most of the mutations do not decrease fitness
or they decrease it by the smallest possible amount of 0.02 (which
means that only after 1 out of the $N$ possible perturbations the
network does not return to the same attractor). We do not show
completely this part of the curves in order to make the distribution
for larger fitness decreases better visible. There are no significant
differences between the curves for different $\mu$. We have already
discussed above that the mean fitness decrease per non-neutral
mutation is larger at later times, when $K$ is larger.

\begin{figure}
\includegraphics*[width=0.4\textwidth]{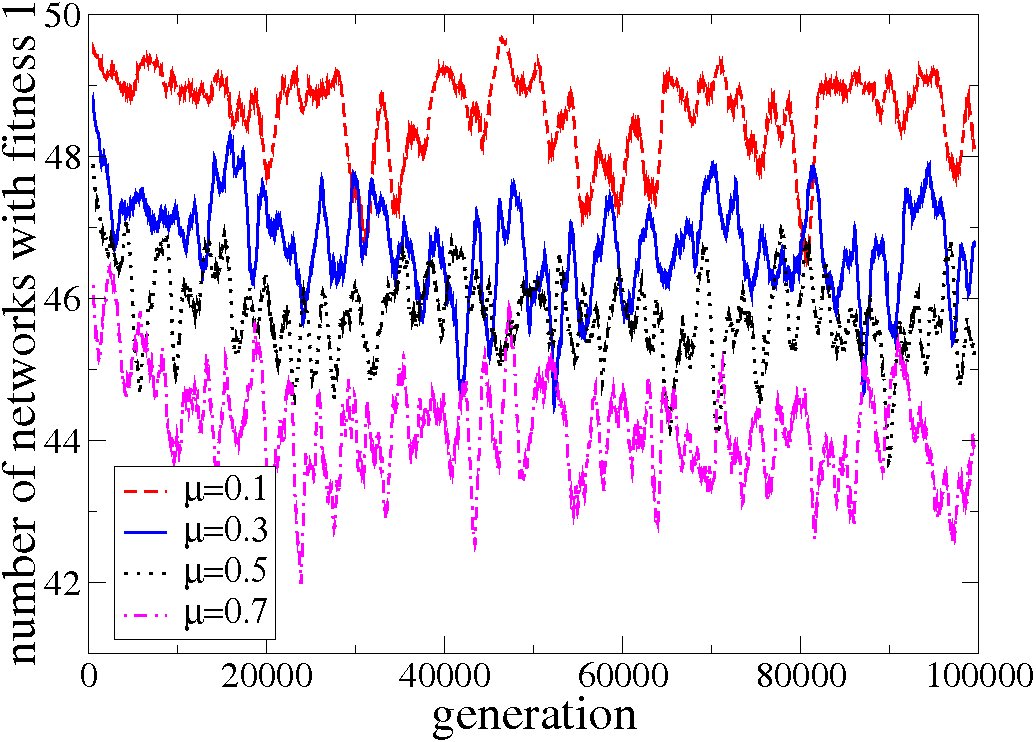}
\caption{
Change of the number of networks with fitness 1 in the population during 
evolution with different mutation rates and high selection pressure
}
\label{DifMFit1nets}
\end{figure}

\begin{figure}
\includegraphics*[width=0.4\textwidth]{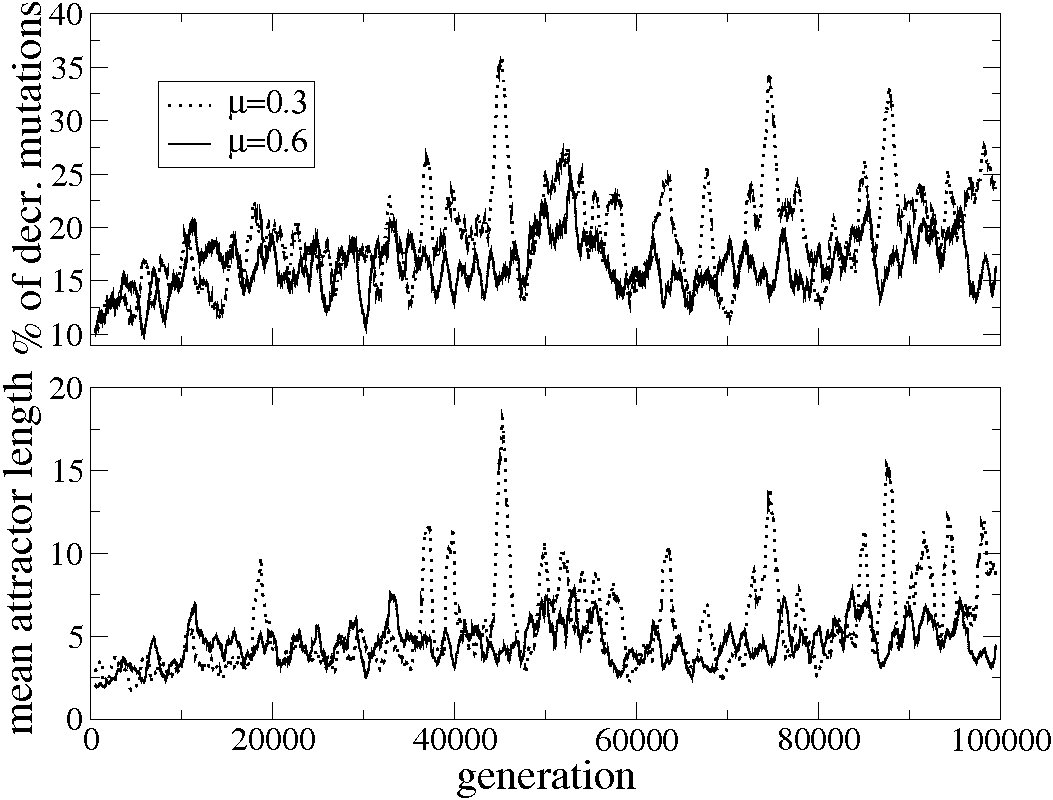}
\caption{ Evolution of the percentage of mutations decreasing fitness 
  (upper graph) and of the mean attractor length in the population (lower
graph) under strong selection and with two different mutation rates. }
\label{attrDecrCorrel}
\end{figure}

While networks evolved with different mutation rates do not differ in
the fitness decrease per (non-neutral) mutation, they do differ in
other respects. 
Figure \ref{DifMFit1nets} shows 
the number of networks with fitness 1 in the population as function of
time for four different mutation rates.
If the probability of a mutation being neutral was the same in all
four cases, the distance of the curves from the value 50 should be
proportional to $\mu$. Due to the large fluctuations, the data cannot
give a clear answer to whether this is the case. 
We therefore evaluated directly the
probability that a mutation  decreases fitness, for mutation rates 
ranging from
$\mu=0.1$ to $\mu=0.7$. These data show a clear trend, with networks
evolved with higher mutation rates being less likely to decrease their
fitness under a mutation. They are more robust to mutations.
Fig.~\ref{attrDecrCorrel}, upper panel, shows the curves obtained with
$\mu=0.3$ and 0.6.
This means that the
networks on the plateau of the fitness landscape, reached by 
evolution under strong selection, have different properties depending
on the mutation rate with which they were evolved. 
This result is not merely due to the fact that populations evolve
slower when the mutation rate is smaller. If this were the case,
the curves obtained with a smaller mutation rate should resemble those
obtained with larger mutation rates at an earlier time. 

There is a correlation between the mean attractor length and the
frequency of neutral mutations, as revealed by the lower graph of
Fig.~\ref{attrDecrCorrel}.  This correlation can be seen most clearly
by comparing the positions of the peaks.  The numerical values of the
correlation between the two curves are 0.892 for $\mu=0.3$ and 0.885
for $\mu=0.6$. These values are not far from the value 1, which would result
if the two curves were proportional to each other.  This means that
networks with longer attractors are more likely to decrease their
fitness under mutations, which is not too surprising.
Conversely, networks that are more robust against mutations
have smaller attractors.

In order to investigate how evolution proceeds when there can be no
slow change in network structure, we also performed simulations with a
different rule, which allows no deletion and addition of links or
changes of functions, but only the redirection of the links. In this
situation, the number of inputs of every node remains 3, and the
distribution of the number of outputs remains Poissonian.  The
population reaches quickly the stationary fitness value, and the
diversity of the population is still large.

\begin{figure}
\includegraphics*[width=0.4\textwidth]{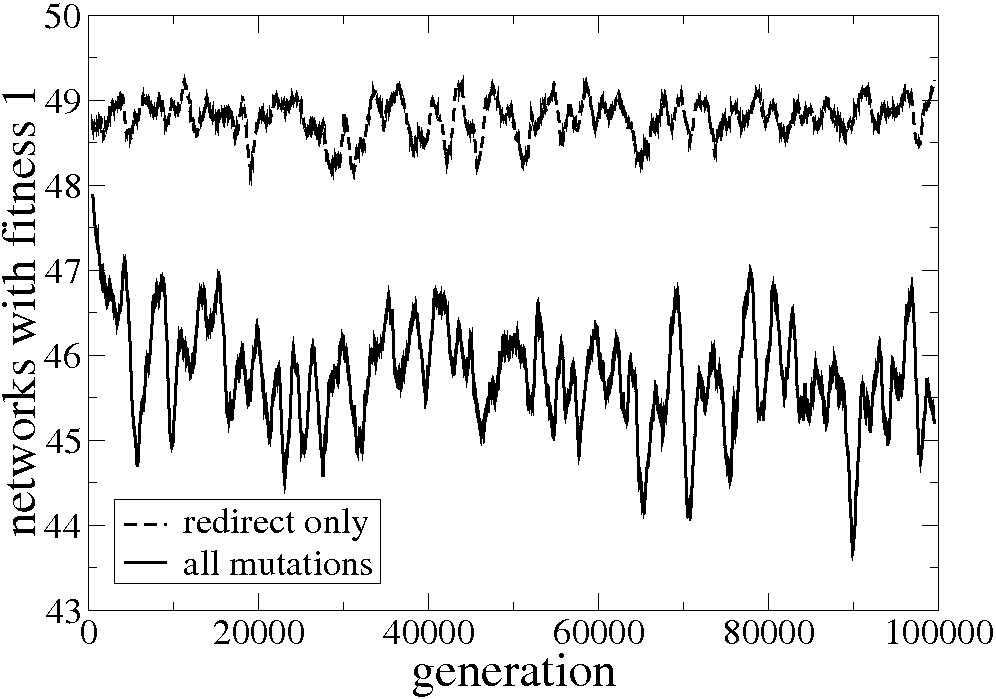}
\caption{ Number of networks with fitness 1 as function of time for
  high selection pressure ($p=100$) and the mutation rate $\mu=0.5$.
  The original model is compared with a model where only the redirection
  of links is allowed.  }
\label{JR+AllFit1nets}
\end{figure}

In Figure \ref{JR+AllFit1nets} the mean number of networks with
fitness 1 in a population obtained with this new mutation rule is
compared to the results obtained with the original rules, for $P=50$
and $\mu=0.5$. The data imply that the mean fitness decrease per
mutation is much larger for the original rule. We have seen
(Fig.~\ref{mutations}) that when the networks are evolved with all
four types of mutations, those mutations redirecting links are
decreasing the fitness less than others. Redirections of links could
also have smaller effect on the fitness of the networks evolved under
the new rule, which would then explain the observed difference.

\begin{figure}
\includegraphics*[width=0.4\textwidth]{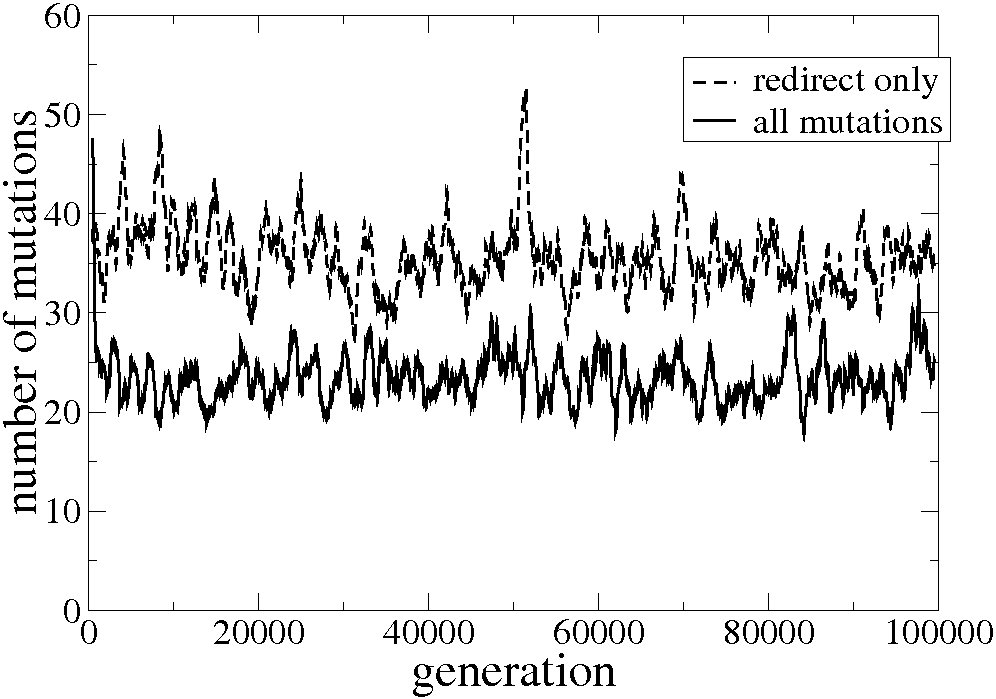}
\caption{
Number of mutations by which
two randomly chosen networks of a population differ, as function of
time, for the two mutation rules. 
The parameters are again $p=100$ and $\mu=0.5$. 
}
\label{DivJR+Allm50}
\end{figure}

Figure \ref{DivJR+Allm50} shows the number of mutations by which
two randomly chosen networks of a population differ, as function of
time, for the two mutation rules. This number is considerably larger when 
links are only rewired. We attribute this result to the larger effective
population size and to the absence of networks with small average $K$ values.

Not surprisingly, we do not find any long-term trend in the mean
fitness of the population with the modified mutation rule. This
confirms our suggestion that the long-term change of the network
structure is responsible for the slow and steady decrease of the
fitness in the original simulations.

Finally, we investigated the influence of the size and of the number
of networks in the population on the properties of evolved populations
by setting $N$ and/or $P$ to 30. We found that the mean fitness of the
population decreases with decreasing $N$, since a node is more likely
to be affected by a mutation when $N$ is smaller.  A decrease of the
population size shows even larger effect on the mean fitness, because
the influence of genetic drift becomes more important compared to the
influence of selection. Earlier in this Section we estimated that the
topological diversity should change approximately
as $P/N$. In our simulations, we find trends that agree with this
assumption. However, due to large fluctuations, we cannot make the
statements of this paragraph more quantitative.


\section{Evolution with finite selection pressure}\label{generalpr}

When selection pressure is finite, the properties of the evolving populations
should be between the two extreme cases studied until now.  When
mutation rates are too high, selection pressures too low, or
population sizes too small, the effect of selection is hardly visible
in the population, the evolution of which is then dominated by drift.

\begin{figure}
\includegraphics*[width=0.4\textwidth]{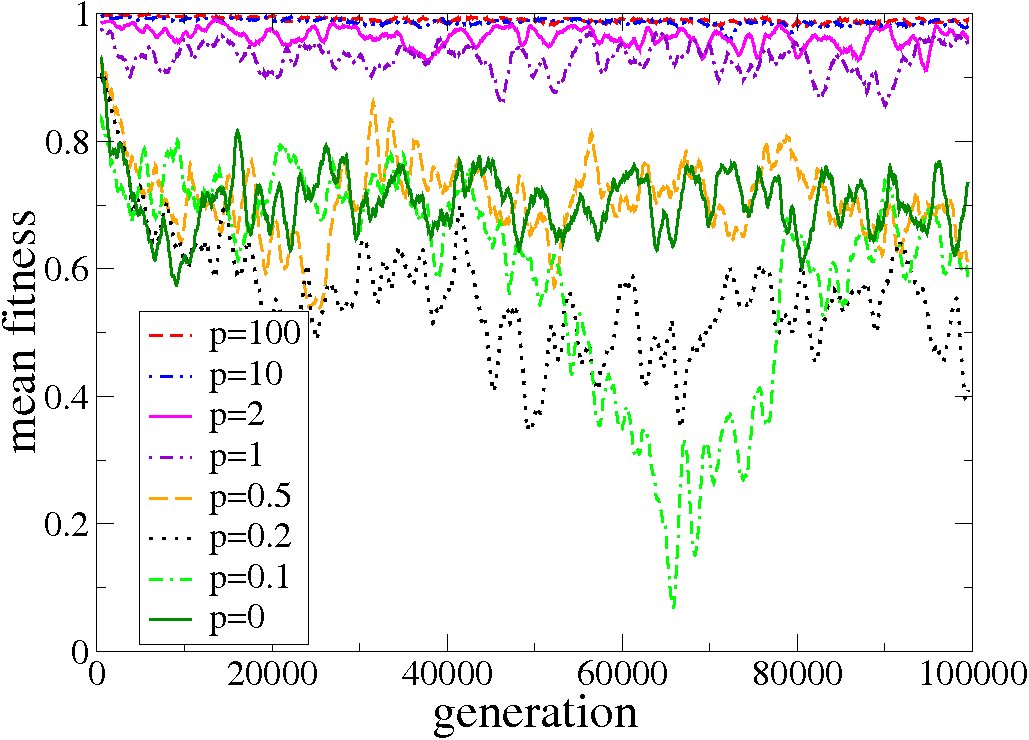}
\caption{
Evolution of the mean fitness for different selection pressures and a mutation
rate $\mu=0.2$.
}
\label{m20fitDiffp}
\end{figure}

\begin{figure}
\includegraphics*[width=0.4\textwidth]{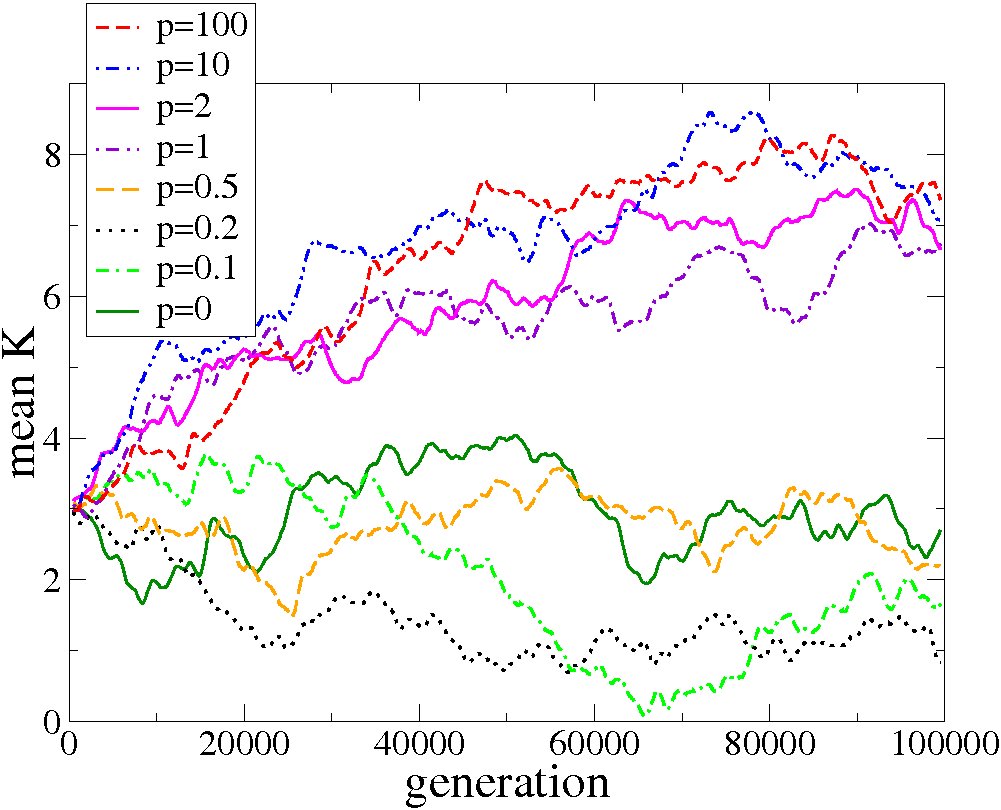}
\caption{
Evolution of the mean number of inputs per node under different selection
pressures and mutation rate 0.2
}
\label{m20Kdiffp}
\end{figure}

Figure \ref{m20fitDiffp} shows the mean fitness of a population with
$N=P=50$ and a mutation rate $\mu=0.2$ as a function of time for different 
values of the selection pressure $p$. One can clearly see that for $p \le 0.5$
the evolution of the fitness resembles that of the system with $p=0$,
which means that drift dominates the evolutionary process. The simulation for
$p=0.1$ accidentally goes through a stage where there are very many
nodes with 0 inputs ( this is correlated with the decrease of the mean $K$ value
that can be seen in Fig.\ref{m20Kdiffp}), resulting in a very low fitness. 

\begin{figure}
\includegraphics*[width=0.4\textwidth]{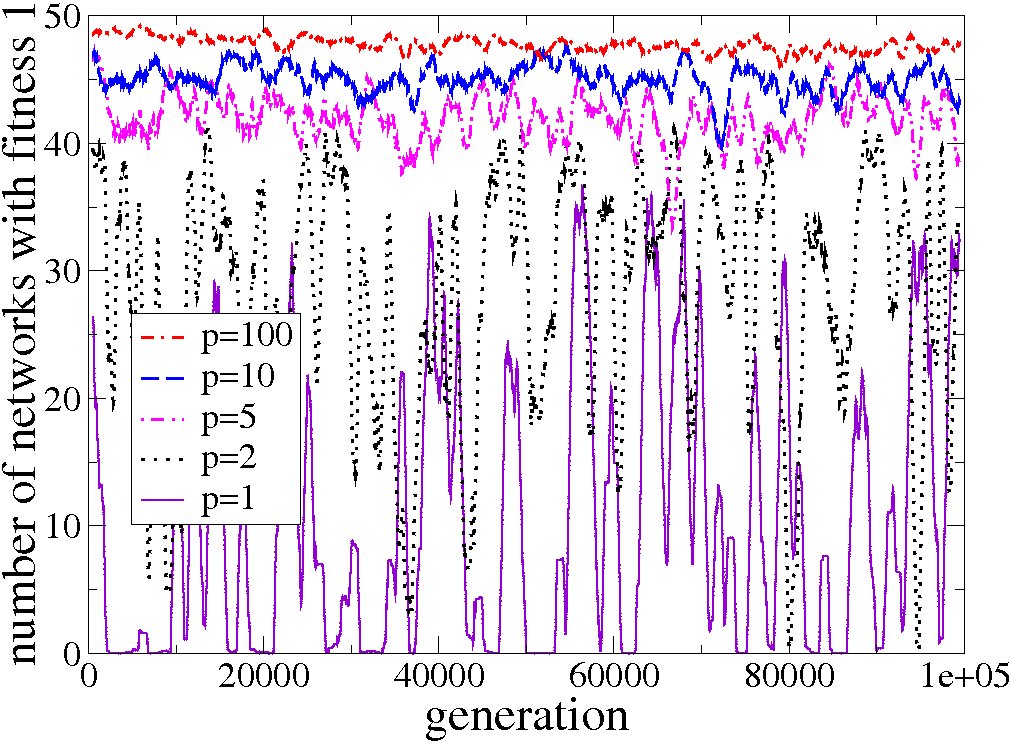}
\caption{
Change of the number of networks with fitness 1 during the evolution under
different selection pressures and mutation rate 0.2
}
\label{fit1m20difP}
\end{figure}

For $p\ge 1$, selection has a clear effect on the fitness. Figure
\ref{fit1m20difP} shows the number of networks with fitness 1 in the
population as a function of time for $p \ge 1$. (For $p \le 0.5$, there
are almost no networks with fitness 1.)  When the selection pressure
is smaller, this number is also smaller.  Just as for the case of very
large selection pressure, the populations show a slow and slight
decrease of the mean fitness with time. This decrease is again
correlated with an increase in the mean connectivity, as shown in
Figure \ref{m20Kdiffp}.

\begin{figure}
\includegraphics*[width=0.4\textwidth]{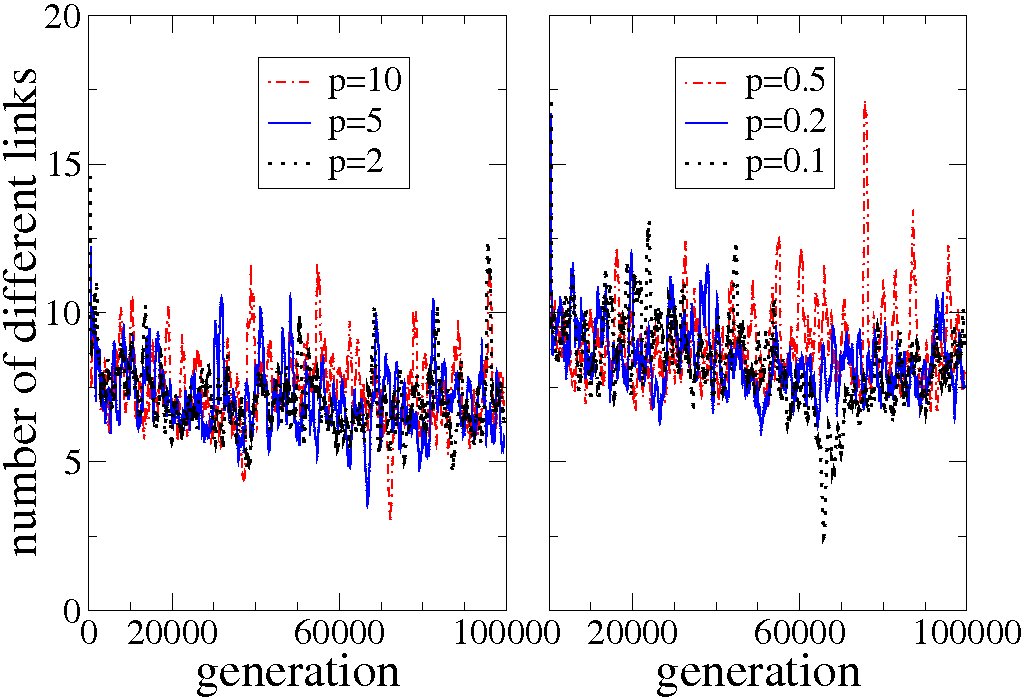}
\caption{
Number of links by which two randomly chosen networks differ
in the population during evolution with mutation rate 0.2 and  under
selection pressures $p\ge 1$ (left) and $p \le 0.5$ (right).
}
 \label{DivMutdiffPm20}
\end{figure}

The topological diversity is again strongly correlated with the number
of inputs per node. In Figure \ref{DivMutdiffPm20} we show 
the number of links by which two
randomly chosen networks of the population differ. This number does
not depend on $K$, but on the effective population size. It has
approximately the same mean value for all selection pressures,
whether they are weak or strong. This is not surprising, as we have
already seen that the effective population sizes do not change much
when the selection pressure is changed from 0 to a very high value.
For weak selection, there are instances where the total  number
of links becomes very small. 

\section{Conclusions}\label{conclusion}

We have investigated the evolution of populations of random Boolean
networks under selection for robustness of the dynamics with respect
to the perturbation of a node. The fitness landscape of such a model
contains a huge plateau of maximum fitness that spans the entire
network space. 

Even in the absence of selection, we found long-term changes in the
network structure. In particular, the distribution of the number of
inputs became broad, leading to a decrease in the mean
fitness. Furthermore, since links are randomly added or deleted during
mutations, the evolutionary process may go through periods where there
are very few links in the networks, which implies that fitness is
particularly low.

When selection is so strong that only networks with the maximum
fitness value 1 can become parents of the next generation, the
evolutionary process is accompanied by a slow increase in the mean
connectivity and a slow decrease in the mean fitness, lasting for
several 10000 generations. In fact, this process was apparently not
fully finished at the end of our long-term simulations.  We ascribe
this long-term trend to the fact that nodes with 0 inputs do not
return to their initial state after a perturbation. They decrease the
fitness of the network, and therefore mutations that add links are
favoured with respect to mutations that remove links. Interestingly,
the mean fitness of the population decreases nevertheless. This
resembles the 'tragedy of commons' \cite{HardinTragedy}, where the
mean fitness of the population decreases, while each individual
strives to obtain maximum fitness. But in contrast to the 'tragedy of
the commons', the fitness of an individual in our model does not
depend on the other individuals. This effect can in our model be
explained by the fact that networks with more links per node are more
likely to decrease their fitness when a mutation is performed.

We found furthermore that populations evolved with higher mutation
rates show a higher robustness against mutations, i.e., they are less
likely to loose fitness under a mutation. This means that even though
all the evolved populations move on the plateau of maximum fitness,
they end up in different regions of network space.  Robustness against
mutations evolves because networks with higher mutational robustness
have more offspring in the next generation. This trend is countered by
the generation of mutationally less robust networks through neutral
mutations. When the effect of mutations becomes more important
(because mutation rates are higher), the equilibrium point between
these two trends moves towards higher mutational robustness.  This
explains why networks evolved under higher mutation rates are more
robust against mutations. We found that higher robustness against
mutations is accompanied by a shorter mean attractor length.
Obviously, higher mutational robustness is also correlated with higher
dynamical robustness, i.e. with higher mean fitness. The above-mentioned
slow decrease (after the initial increase towards the plateau) in the
mean fitness of the population is therefore reflected in a similar
slight decrease in mutational robustness.

Populations evolved at finite selection pressures behave similarly to
those without selection when drift dominates over selection, and they
behave similarly to those with high selection pressure when the effect
of selection dominates over drift. 

Let us now compare the features of our model with those of real gene
regulation networks. The perturbation of the state of a node can be
interpreted as an effect of the omnipresent thermal noise and
stochastic fluctuations of molecular concentrations.  Real gene
regulation networks have to maintain their function under such
perturbations.  This dynamical robustness should be preserved during
evolutionary processes, even when the phenotype of the mutant
individual is different from that of the parent. Indeed, experimental
studies show robustness of cellular networks under mutations. The gene
regulation network of \textit{Escherichia coli}
\cite{isalanRealRewired}, and the \textit{phage} $\lambda$ regulatory
circuitry, \cite{little99real}, preserved their function
under different changes in their structure, implying high
robustness under mutations and high dynamical robustness after the mutations. 
These findings suggest that the fitness landscape of real cellular
networks also contains a huge plateau of high fitness, through which
the networks can move without loss of functionality. 

Other studies of the evolution of Boolean networks show also a
connection between dynamical robustness and mutational robustness. 
In
\cite{wagnerPlasticity,cilibertiRobustTopologies,kaneko07,sevimEvol07},
the authors find that robustness to noise (i.e., to small
perturbations of the state of the network) and robustness to mutations
are highly correlated. In those investigations, mutational robustness
and dynamical robustness evolve together and increase with time. 

The results of our study make it plausible that there is a plateau at
the global maximum of fitness landscape of real genetic regulatory
networks.  We have seen that in such a landscape networks can evolve
without  loss of  dynamical robustness, i.e., of 
functionality. Their evolution is therefore driven by the demand
for achieving high mutational robustness. The evolved networks are
therefore robust to changes in their structure, being in the same time
able to preserve their function under small environmental changes.
When evolution occurs with higher mutation rates, networks can
continue to function only when their mutational robustness is
sufficiently large. Our simulations indicate that an increased
mutational robustness evolves naturally when mutation rates are
higher.  In order to remain evolvable, networks also have to preserve
variability.  This is the case in our simulations, since mutations can
change the phenotype (i.e., the attractors) even after a long time.

We thank Agnes Szejka for useful discussions.

\bibliography{evoliterature}

\end{document}